\documentclass[twocolumn,epjc3]{svjour3}          

\smartqed  

\RequirePackage{graphicx} \journalname{Eur. Phys. J. C}
\def\be{\begin{equation}}
\def\ee{\end{equation}}
\def\bea{\begin{eqnarray}}
\def\eea{\end{eqnarray}}

\begin{document}

\title{A note on Maxwell's equal area law for black hole phase transition}


\author{Shan-Quan Lan\thanksref{addr1}
        \and
        Jie-Xiong Mo\thanksref{addr1}
        \and
        Wen-Biao Liu\thanksref{e1,addr1}
}

\thankstext{e1}{corresponding author \\  e-mail: wbliu@bnu.edu.cn}

\institute{Department of Physics, Institute of Theoretical Physics,
Beijing Normal University, Beijing, 100875, China\label{addr1} }

\date{Received: date / Accepted: date}

\maketitle

\begin{abstract}

The state equation of the charged AdS black hole is reviewed in the $T-r$ plane. Thinking of the phase transition, the $T-S$, $P-V$, $P-\nu$ graphs are plotted and then the equal area law is used in the three cases to get the phase transition point (P,T). The analytical phase transition point relations for P-T of charged AdS black hole has been obtained successfully. By comparing the three results, we find that the equal area law possibly cannot be used directly for $P-\nu$ plane. According to the $T-S$, $P-V$ results, we plot the $P-T-Q$ graph and find that for a highly charged black hole a very low temperature condition is required for the phase transition.

\end{abstract}

\section{Introduction}
\label{introduction}

In 1970s, J.Bekenstein and L.Smarr found out the Smarr relation for a Kerr-Newman black hole and Hawking proved the black hole radiation. Since then the black hole thermodynamics has become an interesting and challenging subject. It turns out that the black hole system is in a precise analogy with the non-gravitational thermodynamic system in nature. A black hole not only has temperature or entropy, but also possesses rich phase structure and admits critical phenomena. Especially, when it comes to the AdS spacetimes, there exists Hawking-Page phase transition between stable large black hole and thermal gas which can be explained as the confinement/deconfinement phase transition of gauge field \cite{1983cmp,1998hepth3131W}. This AdS/CFT correspondence \cite{1999IJTP381113M,1998PhLB428105G,1998AdTMP2253W} can be used to study quark-gluon plasmas and various condensed matter phenomena which makes the AdS black hole thermodynamics more attractive.

In the early time, the charged AdS (RN-AdS) black hole is assigned with thermodynamical variables such as temperature, entropy, charge, electric potential, but not the volume and pressure. An attempt to introduce the pressure was carried out in Refs.\cite{1999PhRvD60f4018C,1999PhRvD60j4026C}. They rearranged the characters of thermodynamical variables: $1/T$ was identified with $P$, $r_{+}$ was identified with $V$, $Q$ was identified with $T$. This approach is not uniquely defined and mismatches intensive and extensive thermodynamical variables \cite{2012JHEP07033K}. Recently, treating the variation of cosmological constant $\Lambda$ as pressure has attained increasing attention \cite{2012JHEP07033K,2000CQGra17399C,2009CQGra26s5011K,2011CQGra28l5020D,2011CQGra28w5017D,2011PhRvD84b4037C,2012PhRvD86d4011L,2014CQGra31d2001A,2014arXiv14117850D,2014PhRvD90d4057W,2013arXiv13053379S,2014EPJC743052Z,2014PhRvD89h4057M}. This interpretation is much more physically sound and avoids the confusion among intensive and extensive variables. What's more, when $\Lambda$ is identified with pressure and the black hole mass $M$ is identified with enthalpy rather than internal energy\cite{2009CQGra26s5011K}, the first law is consistent with the Smarr relation\cite{1997hepth2087R,200537643,2010090404}.

With this interpretation of the cosmological constant, many works on AdS black hole phase transition were carried out \cite{2013PhRvD88j1502A,2014arXiv1412.5028S,2014arXiv1412.3880C,2014arXiv1411.3554Z,2014arXiv1410.0352H,2014PhLB735256L,2014PhLB736214X,2014JHEP09179L,2014PhRvD90d4063Z}. In Ref.\cite{2012JHEP07033K}, the authors showed that the graph plotted by state equation of AdS black hole for fixed $T$ in the $P-r$ or $P-V$ plane is a reminiscence of Van der Waals (VDW) system. There is an oscillating part of the graph (for $T<T_{c}$) which denotes a phase transition. The oscillating part needs to be replaced by an isobar such that the areas above and below it are equal to one another. This treatment is called the Maxwell's equal area law. They also showed that the graph of the Gibbs free energy (during the phase transition progress) which possesses a characteristic swallow tail is also a reminiscence of VDW system. What's more, they studied the critical exponents and showed that the results coincide with those of the VDW system. That is why we use the equal area law which is valid for the VDW system to address the phase transition of RN-AdS black hole.

In our knowledge, there are two ways to obtain the phase diagram or the coexistence line of AdS black hole: (1)using the Maxwell's equal area law, (2)analyzing the characteristic swallow tail behavior of the Gibbs free energy. In this paper, we will use the equal area law to obtain the coexistence line ($P-T$) and the coexistence surface ($P-T-Q$) of RN-AdS black hole. Especially, we use the method which was used in $T-S$ plane in Ref.\cite{2013arXiv13053379S} to explicitly argue the viewpoint raised in Ref.\cite{2014shaowen} that the equal area law can be only used in the $P-V$, $T-S$ and $\phi-Q$ planes while not in the $T-r$ or $P-\nu$ plane though for these cases there is also an oscillating behavior below the critical point. Moreover, we obtained the analytical phase transition point relation P-T-Q of RN-AdS black hole from $P-V$ plane.

This paper is organized as follows. In Sec.\ref{rnadsbh}, we review the thermodynamical state equation of the RN-AdS black hole in $T-r$ plane. We use the critical point to rescale the state equation and have a brief analyse of the oscillating part. In Sec.\ref{ealpt}, we use the equal area law in $T-S$, $P-V$ and $P-\nu$ planes to obtain the analytical phase transition point relation P-T-Q and discuss the equal area law, Smarr relation and the first law. In Sec.\ref{conclusion}, we discuss and conclude some results.

\section{Thermodynamic state equation of charged AdS black hole and the system's critical point}
\label{rnadsbh}

The line element of the RN-AdS metric is given as
\begin{equation}
ds^{2}=-Fdt^{2}+\frac{dr^{2}}{F}+d\Omega_{2}^{2},
\end{equation}
where $d\Omega_{2}^{2}$ stands for the standard element on $S^{2}$ and the function $F$ is given by
\begin{equation}
F=1-\frac{2M}{r}+\frac{Q^{2}}{r^{2}}+\frac{r^{2}}{l^{2}}.
\end{equation}
The parameter $M$ represents the ADM mass of the black hole, Q represents the total charge and the AdS curvature radius $l$ is related to the cosmological constant as $\Lambda=-3/l^2$.

The black hole radius, same as the event horizon position, is determined as the largest root of $F(r_{+})=0$. The black hole temperature $T$ is given by
\begin{eqnarray}
T&=&\frac{\kappa}{2\pi}=\frac{1}{2\pi}(-\frac{1}{2}\sqrt{\frac{g^{11}}{-\hat{g}_{00}}}\hat{g}_{00,1}) \nonumber\\
&=&\frac{1}{4\pi{r_{+}}}(1+\frac{3r_{+}^{2}}{l^{2}}-\frac{Q^{2}}{r_{+}^{2}}).
\end{eqnarray} \\
By setting $T=0$, we can get the extremal RN-AdS black hole as \cite{2013arXiv13053379S}
\begin{eqnarray}
r_{0}^{2}=\frac{l^{2}}{6}(\sqrt{1+\frac{12Q^{2}}{l^{2}}}-1), M_{0}=\frac{r_{0}}{3}(2+\sqrt{1+\frac{12Q^{2}}{l^{2}}}).\nonumber\\
\end{eqnarray}

In fact Eq.(3) is the thermodynamic state equation of the black hole and we will frequently use it when we argue about the equal area law. Now we are going to find out the critical point and some state information of the RN-AdS black hole from $T-r_{+}$ plane. The $T-r_{+}$ plane is preferred in Ref.\cite{2013arXiv13053379S} because Eq.(3) here is original thermodynamic state equation regardless of thermodynamic quantities $P$ and $V$. We treat $Q$ as a constant, then $T$ is a function of $r_{+}$. The critical point occurs when $T(r_{+})$ has an inflection point
\begin{equation}
\frac{\partial{T}}{\partial{r_{+}}}=0 ,\,\, \frac{\partial^{2}{T}}{\partial{r_{+}^{2}}}=0.
\end{equation}
From the above two equations, we get the critical values ( we replace $\frac{3}{8\pi{l}^{2}}$ by $P$, it's just for convenience here, but we will see in Ref.\cite{2009CQGra26s5011K} that $P$ is treated as pressure of the AdS black hole system )
\begin{eqnarray}
r_{c}=\sqrt{6}Q , \frac{3}{8\pi{l_{c}}^{2}}=\frac{1}{96\pi{Q}^{2}}\equiv{P_{c}}, T_{c}=\frac{\sqrt{6}}{18\pi{Q}}.
\end{eqnarray}

\begin{figure}
\begin{center}
\includegraphics[width=0.48\textwidth]{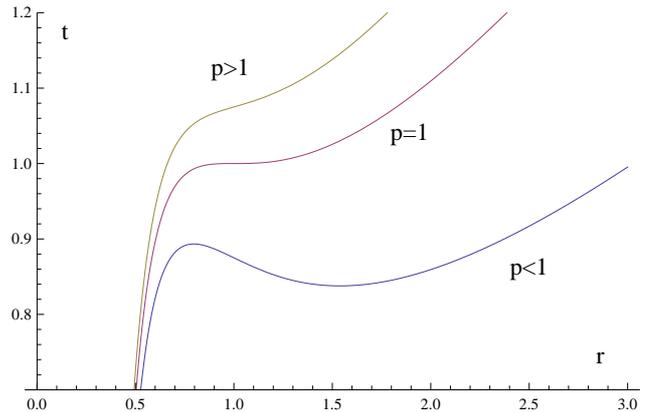}
\caption{The rescaled $t-r$ thermodynamical state graph of RN-AdS black hole for different $p$. The critical point is at $p=1,r=1,t=1$ which is an inflection point.}  \label{ttr}
\end{center}
\end{figure}

From Eqs.(6) we can see $\frac{P_{c}r_{c}}{T_{c}}=\frac{3}{16}$ , so we can rescale Eq.(3) to get rid of $Q$ by defining \cite{2013arXiv13053379S}
\begin{equation}
r_{+}=rr_{c} ,\,\,\,\,\, P=pP_{c} ,\,\,\,\,\, T=tT_{c}.
\end{equation}
Then the rescaled equation of $T$ becomes
\begin{equation}
t=\frac{3}{4}(\frac{1}{r}+\frac{rp}{2}-\frac{1}{6r^{3}}).
\end{equation}
The $t-r$ graphs for different $p$ are plotted in Fig.\ref{ttr}.

For $p=1$ , there is an inflection point at $t=1,r=1$ which is just the critical point. For $p>1$ which is beyond the critical point, the temperature $t$ increases as radius $r$ increases. It's a thermally stable ``gas" phase black hole (here the ``gas" means that the black hole radius changes largely as the temperature changes just like the gas does, and the ``liquid" means that the black hole radius almost doesn't change as the temperature changes just like the liquid does). For $p<1$ which is below the critical point, the $t-r$ graph is oscillating. There are one maximum point at $(r_{max},t_{max})$ and one minimum point at $(r_{min},t_{min})$. For $r<r_{max}$ , the temperature $t$ increases as radius $r$ increases, it's a small, thermally stable, near-extremal black hole. For $r_{max}<r<r_{min}$, the temperature $t$ decreases as radius $r$ increases, it's a thermally unstable black hole. For $r_{min}<r$, the temperature $t$ increases as radius $r$ increases, it's a large, thermally stable black hole. Inspired by the VDW gas-fluid $P-V$ graph, we can replace the oscillating part of the isopiestic for $p<1$ by an isobar. For the VDW $P-V$ graph, the isobar should make sure that the areas above and below it are equal to one another which is called the equal area law. One may say we can use the equal area law here too. In fact, in Ref.\cite{2013arXiv13053379S}, they applied the method for $P-\nu$ graph. But we find out that the equal area law can't be used here. We will show that the oscillating part should be replaced by an isobar but not the equal area isobar later. Now, we suppose that the oscillating part is replaced by an isobar of which the left hand side is $r_{l}$ and the right hand side is $r_{g}$. Just like the VDW case, from right to left, for $r>r_{g}$ the black hole system is in its "gas" phase, for $r_{l}<r<r_{g}$ it goes through a phase transition from ``gas" phase to ``liquid" phase, for $r<r_{l}$ it is in "liquid" phase.

\section{Equal area law and phase transition}
\label{ealpt}

It is firstly suggested in Ref.\cite{2009CQGra26s5011K} that we may regard the cosmological constant as a variable and treat it as a dynamical pressure of black hole
\begin{equation}
p=-\frac{1}{8\pi}\Lambda=\frac{3}{8\pi{l^{2}}}.
\end{equation}
With this interpretation, the black hole mass is identified with enthalpy rather than internal energy. Recently, there is a great interest on studying the phase transition of the RN-AdS black hole system in this extended phase space. Based on Ref.\cite{2013arXiv13053379S}, we will use the equal area law in $T-S$, $P-V$ and $P-\nu$ planes to investigate the phase transition in detail.

\subsection{Equal area law in T-S plane}
\label{ts}

\begin{figure}
\begin{center}
\includegraphics[width=0.48\textwidth]{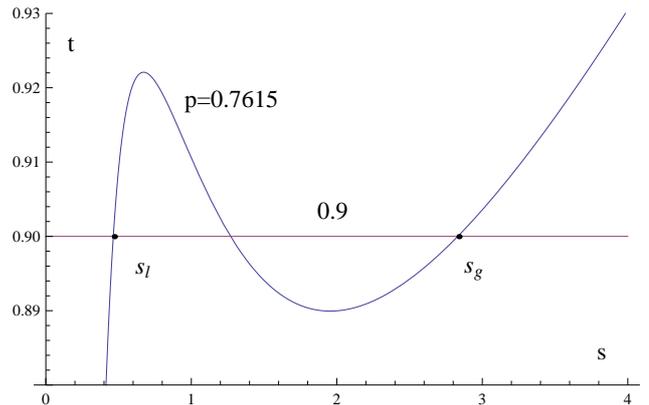}
\caption{The rescaled $t-s$ graph of RN-AdS black hole for $p=0.7615$. The oscillating areas above and below the straight line are equal to one another. The equal area law tells us the phase transition point is at $p=0.7615,t=0.9$. } \label{tts}
\end{center}
\end{figure}

The entropy of the black hole system is
\begin{equation}
S=\frac{A}{4},A=4\pi{r_{+}^{2}}.
\end{equation}
We can rewrite the thermodynamic state equation as
\begin{equation}
T=\frac{1}{4\sqrt{\pi{S}}}(1+8PS-\frac{\pi{Q^{2}}}{S}).
\end{equation}
At the critical point, the entropy is $S_{c}=\pi{r_{c}^{2}}=6\pi{Q^{2}}$. So just as the above section, we can rescale $S=sS_{c}$. Together with the rescaled $T$ and $P$, the state equation can be written as
\begin{equation}
t=\frac{3}{8}(p\sqrt{s}+\frac{2}{\sqrt{s}}-\frac{1}{3s^{\frac{3}{2}}}).
\end{equation}
To show the phase transition, $t,p$ should be smaller than the critical point which means $t<1$ and $p<1$. We plot the graph $t-s$ for $p=0.7615$ in Fig.\ref{tts}.

The oscillating part should be replaced by an isobar which satisfies that the areas above and below it are equal to one another. We suppose the isobar is $t=t^{*}$, so the left cross point is $s_{l}$ denoting the ``liquid" phase entropy and the right cross point is $s_{g}$ denoting the ``gas" phase entropy. Then the equal area law is manifested as
\begin{equation}
t^{*}(s_{g}-s_{l})=\int_{s_{l}}^{s_{g}}t(s)ds,
\end{equation}
so we obtain
\begin{eqnarray}
&\,&t^{*}=\frac{1}{\sqrt{s_{g}}+\sqrt{s_{l}}}(\frac{3}{2}-\frac{1}{4\sqrt{s_{l}s_{g}}} \nonumber\\
&\,\,\,\,\,&+\frac{p^{*}}{4}(s_{l}+s_{g}+\sqrt{s_{l}s_{g}})).
\end{eqnarray}

Our purpose is to find out the coexistence phase transition line for $(t,p)$, thus we just need to solve Eqs.(12) and (14). For Eq.(12), from Fig.\ref{tts}, the left and right cross points give us two equations. Using $x=\sqrt{s_{l}},y=\sqrt{s_{g}}$, together with Eq.(14) we have
\begin{equation}
t^{*}=\frac{3}{8}(p^{*}x+\frac{2}{x}-\frac{1}{3x^{3}}),
\end{equation}
\begin{equation}
t^{*}=\frac{3}{8}(p^{*}y+\frac{2}{y}-\frac{1}{3y^{3}}),
\end{equation}
\begin{equation}
t^{*}=\frac{1}{x+y}(\frac{3}{2}-\frac{1}{4xy}+\frac{p^{*}}{4}(x^{2}+y^{2}+xy)).
\end{equation}
The equations can be solved straightforwardly. However, it is difficult and complicated. Now we will calculate it in a simpler way.

$Eq.(15) - Eq.(16)=0$ will give
\begin{equation}
x^{2}+y^{2}=-3p^{*}x^{3}y^{3}+6x^{2}y^{2}-xy.
\end{equation}
$ 2*Eq.(17)=Eq.(15)+Eq.(16)$ will give
\begin{eqnarray}
&\,&p^{*}(x^{2}+y^{2})x^{3}y^{3}-6(x^{2}+y^{2})x^{2}y^{2}+(x^{2}+y^{2})^{2} \nonumber\\
&\,&+(x^{2}+y^{2})xy=2p^{*}x^{4}y^{4}-12x^{3}y^{3}+2x^{2}y^{2}.
\end{eqnarray}
Then putting Eq.(18) into Eq.(19) to get rid of $x^{2}+y^{2}$ and making a replacement $z=xy$, we have
\begin{equation}
{p^{*}}^{2}z^{4}-2p^{*}z^{3}+2z-1=0=z^{2}(p^{*}z-1)^{2}-(z-1)^{2},
\end{equation}
so we obtain
\begin{eqnarray}
z_{1,2}=\pm\frac{1}{\sqrt{p^{*}}},\,\,\, z_{3,4}=\frac{1}{p^{*}}(1\pm\sqrt{1-p^{*}}),
\end{eqnarray}
where $z_{2}<0$ is meaningless and $z_{3,4}$ are also unreasonable as $s_{l}=s_{g}$. So we only have $z=\frac{1}{\sqrt{p^{*}}}$. Together with Eq.(18), we get
\begin{equation}
s_{g,l}=\frac{1}{2p^{*}}(\sqrt{3-\sqrt{p^{*}}}\pm\sqrt{3-3\sqrt{p^{*}}})^{2},
\end{equation}
\begin{equation}
t^{*}=\sqrt{p^{*}(3-\sqrt{p^{*}})/2}.
\end{equation}
This is the coexistence line of ``gas" and ``liquid" phases for the black hole system. We can also rewrite it as
\begin{equation}
p^{*}=[1-2cos(\frac{arccos(1-{t^{*}}^{2})+\pi}{3})]^{2}.
\end{equation}
At last, we can rescale back $P=pP_{c},T=tT_{c}$ to get the phase transition phase as
\begin{equation}
T=\sqrt{\frac{8P(3-\sqrt{96\pi{Q^{2}}P)}}{9\pi}},
\end{equation}
with $T<T_{c}=\frac{\sqrt{6}}{18\pi{Q}}$ and $P<P_{c}=\frac{1}{96\pi{Q^{2}}}$. For $P>P_{c}$, the black hole only exists in its ``gas" phase.

\subsection{Equal area law in P-V plane}
\label{pv}

\begin{figure}
\begin{center}
\includegraphics[width=0.48\textwidth]{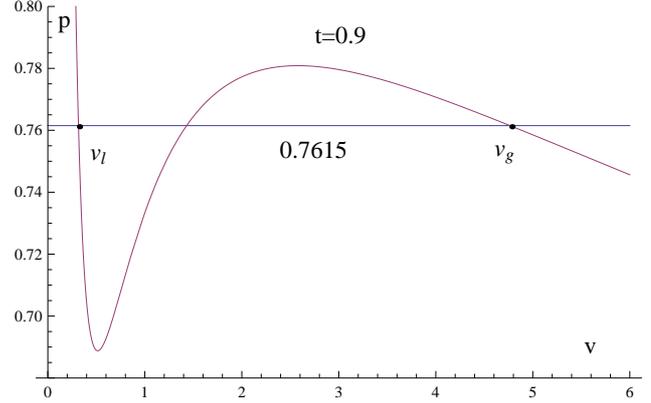}
\caption{The rescaled $p-v$ graph of RNAdS black hole for $t=0.9$. The oscillating areas above and below the straight line are equal to one another. The equal area law tells us the phase transition point is at $p=0.7615,t=0.9$. } \label{tpv1}
\end{center}
\end{figure}

The volume of the black hole is
\begin{equation}
V=\frac{4}{3}\pi{r_{+}}^{3}.
\end{equation}
Thus we can rewrite the thermodynamical state equation as
\begin{equation}
P=\frac{T}{2}(\frac{4\pi}{3V})^{\frac{1}{3}}-\frac{1}{8\pi}(\frac{4\pi}{3V})^{\frac{2}{3}}+\frac{Q^{2}}{8\pi}(\frac{4\pi}{3V})^{\frac{4}{3}}.
\end{equation}
At critical point, $V_{c}=\frac{4}{3}\pi{r_{c}}^{3}=8\sqrt{6}\pi{Q}^{3}$. Thus we can use the same rescaling method $V=vV_{c}$ together with $T,P$ to rewrite the above state equation
\begin{equation}
p=\frac{8t}{3}v^{-\frac{1}{3}}-2v^{-\frac{2}{3}}+\frac{1}{3}v^{-\frac{4}{3}}.
\end{equation}
We plot the $p(v)$ graph in Fig.\ref{tpv1} for $t=0.9$.

The oscillating part should be replaced by an isobar which satisfies that the areas above and below the isobar are equal to one another. We suppose the isobar is $p=p^{*}$, the left cross point is $v_{l}$ denoting the ``liquid" phase volume and the right cross point is $v_{g}$ denoting the ``gas" phase volume. Then the equal area law is manifested as
\begin{equation}
p^{*}(v_{g}-v_{l})=\int_{v_{l}}^{v_{g}}p(v)dv,
\end{equation}
so we obtain
\begin{eqnarray}
&\,&p^{*}=\frac{1}{v_{g}-v_{l}}[4t^{*}({v_{g}}^{\frac{2}{3}}-{v_{l}}^{\frac{2}{3}})-6({v_{g}}^{\frac{1}{3}}-{v_{l}}^{\frac{1}{3}})\nonumber\\
&\,&-({v_{g}}^{-\frac{1}{3}}-{v_{l}}^{-\frac{1}{3}})].
\end{eqnarray}
Using $x^{3}=v_{l},y^{3}=v_{g}$, from Eqs.(28)and (30),we have
\begin{equation}
p^{*}=\frac{8t^{*}}{3x}-\frac{2}{x^{2}}+\frac{1}{3x^{4}},
\end{equation}
\begin{equation}
p^{*}=\frac{8t^{*}}{3y}-\frac{2}{y^{2}}+\frac{1}{3y^{4}},
\end{equation}
\begin{equation}
p^{*}=\frac{4t^{*}(x+y)-6+\frac{1}{xy}}{x^{2}+xy+y^{2}}.
\end{equation}

$Eq.(31) - Eq.(32)=0$ will give
\begin{equation}
8t^{*}x^{3}y^{3}-6(x+y)x^{2}y^{2}+(x^{2}+y^{2})(x+y)=0.
\end{equation}
$ 2*Eq.(33)=Eq.(31)+Eq.(32)$ will give
\begin{eqnarray}
&\,&24t^{*}(x+y)x^{5}y^{5}-36x^{4}y^{4}+6x^{3}y^{3}=(x^{2}+xy+y^{2})[\nonumber\\
&\,&8t^{*}(x+y)x^{3}y^{3}-6(x^{2}+y^{2})x^{2}y^{2}+x^{4}+y^{4}].
\end{eqnarray}
Eqs.(34) and (35) can be simplified as
\begin{equation}
2t^{*}x^{2}y^{2}=x+y,
\end{equation}
\begin{equation}
2{t^{*}}^{2}x^{3}y^{3}-3xy+1=0.
\end{equation}
According to Eq.(37), we get
\begin{eqnarray}
&\,&xy=-\frac{\sqrt{2}}{t^{*}}cos(\frac{\theta}{3}),\,\,\frac{\sqrt{2}}{t^{*}}cos(\frac{\theta+\pi}{3}),\,\,\frac{\sqrt{2}}{t^{*}}cos(\frac{\pi-\theta}{3}),\nonumber\\
&\,&cos(\theta)=\frac{\sqrt{2}}{2}t^{*}.
\end{eqnarray}
Only the last one is correct as $xy$ should be larger than 1. Putting it back into Eq.(36) and using $\varphi=\frac{\pi-\theta}{3}$, we have
\begin{eqnarray}
x&=&\frac{2}{t^{*}}cos^{2}\varphi-\sqrt{\frac{4}{{t^{*}}^{2}}cos^{4}\varphi-\frac{\sqrt{2}}{t^{*}}cos\varphi}, \nonumber\\
y&=&\frac{2}{t^{*}}cos^{2}\varphi+\sqrt{\frac{4}{{t^{*}}^{2}}cos^{4}\varphi-\frac{\sqrt{2}}{t^{*}}cos\varphi}, \nonumber\\
\varphi&=&\frac{\pi-\theta}{3}, cos\theta=\frac{\sqrt{2}}{2}t^{*}.
\end{eqnarray}
So from Eq.(32) we have
\begin{eqnarray}
p^{*}&=&\frac{8t^{*}}{3y}-\frac{2}{y^{2}}+\frac{1}{3y^{4}}\nonumber\\
&=&\frac{8t^{*}}{3}(\sqrt{2}cos\varphi-\sqrt{2cos^{2}\varphi-\frac{t^{*}}{\sqrt{2}cos\varphi}})\nonumber\\
&\,\,&-2(\sqrt{2}cos\varphi-\sqrt{2cos^{2}\varphi-\frac{t^{*}}{\sqrt{2}cos\varphi}})^{2}\nonumber\\
&\,\,&+\frac{1}{3}(\sqrt{2}cos\varphi-\sqrt{2cos^{2}\varphi-\frac{t^{*}}{\sqrt{2}cos\varphi}})^{4}\nonumber\\
&=&16(cos\frac{\theta}{3}cos\frac{\pi+\theta}{3})^{2}\nonumber\\
&=&(1-2cos\frac{2(\pi-arccos\frac{t^{*}}{\sqrt{2}})}{3})^{2}\nonumber\\
&=&(1-2cos\frac{arccos(1-{t^{*}}^{2})+\pi}{3})^{2},
\end{eqnarray}
where $0\leq{t^{*}}\leq1$ is just the physical cases that we are interested. Now, we have
\begin{equation}
t^{*}=\sqrt{p^{*}(3-\sqrt{p^{*}})/2}.
\end{equation}
We find that this result is just same with Eq.(23). This means the equal area law is identical in $T-S$ plane and $P-V$ plane. So far we get the analytical result of $p(t)$ or $P(T,Q)$ (rescale back $p(t)$ ) for the coexistence line using $P-V$ equal area law.

\subsection{Equal area law in $P-\nu$ plane}  \label{third}

\begin{figure}
\begin{center}
\includegraphics[width=0.48\textwidth]{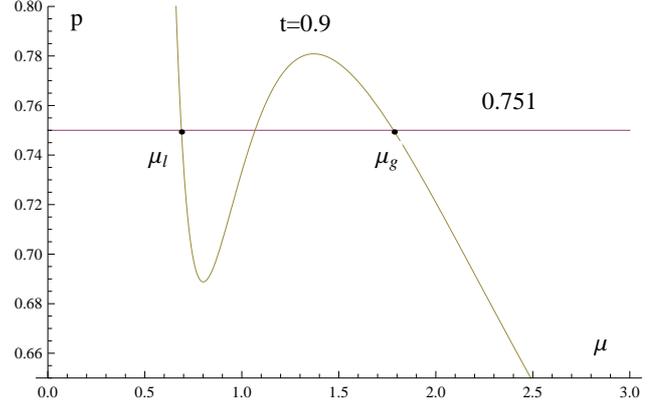}
\caption{The rescaled $p-\mu$ graph of RNAdS black hole for $t=0.9$. The oscillating areas above and below the straight line are equal to one another. The equal area law tells us the phase transition point is at $p=0.751,t=0.9$. }  \label{tpv2}
\end{center}
\end{figure}

In Refs.\cite{2012JHEP07033K} and \cite{2013arXiv13053379S}, the RN-AdS black hole's thermodynamical state equation is compared with the VDW equation and the specific volume is identified
\begin{equation}
\nu=2l_{p}^{2}r_{+},
\end{equation}
where $l_{p}$ denotes the Planck length. In this subsection, we will use the equal area law in $P-\nu$ plane to get the coexistence line $p(t)$. By setting $l_{p}=1$, the state equation can be written as
\begin{equation}
P=\frac{T}{\nu}-\frac{1}{2\pi{\nu}^{2}}+\frac{2Q^{2}}{\pi{\nu}^{4}}.
\end{equation}
At the critical point, $\nu_{c}=2\sqrt{6}Q$. We rescale it by $\nu=\mu\nu_{c}$. Together with rescaled $T,P$, we have
\begin{equation}
p=\frac{8t}{3\mu}-\frac{2}{\mu^{2}}+\frac{1}{3\mu^{4}}.
\end{equation}
We plot the graph in Fig.\ref{tpv2} for $t=0.9$.

The oscillating part should be replaced by an isobar which satisfies that the areas above and below the isobar are equal to one another. We suppose the isobar is $p=p^{*}$, the left cross point is $\mu_{l}$ denoting the ``liquid" phase volume and the right cross point is $\mu_{g}$ denoting the ``gas" phase volume. Then the equal area law is manifested as
\begin{equation}
p^{*}(\mu_{g}-\mu_{l})=\int_{\mu_{l}}^{\mu_{g}}p(\mu)d\mu,
\end{equation}
so we obtain
\begin{eqnarray}
p^{*}&=&\frac{1}{\mu_{g}-\mu_{l}}(\frac{8t^{*}}{3}ln(\frac{\mu_{g}}{\mu_{l}})+\frac{2}{\mu_{g}}\nonumber\\
&\,&-\frac{2}{\mu_{l}}-\frac{1}{9\mu_{g}^{3}}+\frac{1}{9\mu_{l}^{3}}).
\end{eqnarray}
Using $x=\frac{1}{\mu_{l}},y=\frac{1}{\mu_{g}}$, from Eqs.(44) and (46) we have
\begin{equation}
p^{*}=\frac{8t^{*}}{3}x-2x^{2}+\frac{1}{3}x^{4},
\end{equation}
\begin{equation}
p^{*}=\frac{8t^{*}}{3}y-2y^{2}+\frac{1}{3}y^{4},
\end{equation}
\begin{equation}
p^{*}=\frac{xy}{x-y}(\frac{8t^{*}}{3}ln\frac{x}{y}-2(x-y)+\frac{1}{9}(x^{3}-y^{3})).
\end{equation}
Using the straightforward way, an implicit function of $p^{*}(t^{*})$ can be obtained.
\begin{eqnarray}
&&x_{1}=\sqrt{1+\frac{\Delta}{4}}-\sqrt{2-\frac{\Delta}{4}-\frac{4t^{*}}{\sqrt{4+\Delta}}},\nonumber\\
&&x_{2}=\sqrt{1+\frac{\Delta}{4}}+\sqrt{2-\frac{\Delta}{4}-\frac{4t^{*}}{\sqrt{4+\Delta}}},\nonumber\\
&&x_{3}=-\sqrt{1+\frac{\Delta}{4}}-\sqrt{2-\frac{\Delta}{4}+\frac{4t^{*}}{\sqrt{4+\Delta}}},\nonumber\\
&&x_{4}=-\sqrt{1+\frac{\Delta}{4}}+\sqrt{2-\frac{\Delta}{4}+\frac{4t^{*}}{\sqrt{4+\Delta}}},\nonumber\\
&&\Delta{\equiv}\frac{2(1-p^{*})}{A^{\frac{1}{3}}}+2A^{\frac{1}{3}}, \nonumber\\
&&A\equiv4{t^{*}}^{2}-3p^{*}-1+\sqrt{(p^{*}-1)^{3}+(1+3p^{*}-4{t^{*}}^{2})^{2}}.\nonumber\\
\end{eqnarray}
For $1<x$ and $0<y<1$, we identify that $x=x_{2}$ and $y=x_{4}$. Putting them back to Eq.(49), we get an implicit function of $p^{*}=p^{*}(t^{*})$ as
\begin{eqnarray}
&\,&p^{*}=\frac{x_{2}x_{4}}{x_{2}-x_{4}}[\frac{8t^{*}}{3}ln\frac{x_{2}}{x_{4}}\nonumber\\
&\,&-2(x_{2}-x_{4})+\frac{1}{9}({x_{2}}^{3}-{x_{4}}^{3})].
\end{eqnarray}

\subsection{The Smarr relation and the first law}  \label{tflpt}

In the above three subsections, we have got three phase transition coexistence lines for RN-AdS black hole while only two lines are identical. However a thermodynamical system should have only one real phase transition coexistence line, which means the same phase transition point at $t=0.9$ in Fig.\ref{tts} and Fig.\ref{tpv1} and Fig.\ref{tpv2}. Unfortunately, the one obtained from $P-\nu$ plane is not identical to the other two cases. In the black hole phase transition research, the $P-\nu$ possibly cannot be used to identify the transition point by the equal area law directly.

\begin{figure}
\begin{center}
\includegraphics[width=0.40\textwidth]{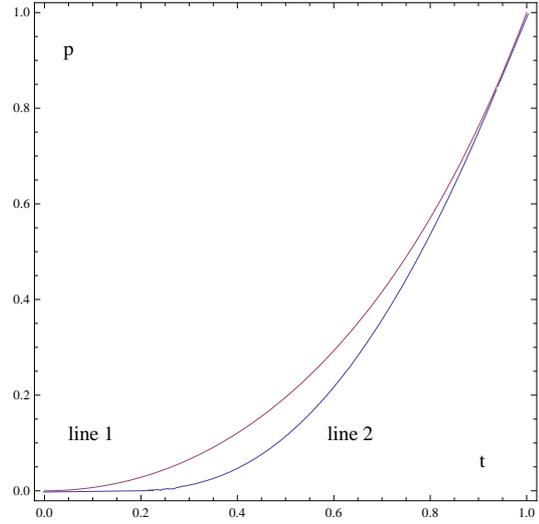}
\caption{The phase transition coexistence lines (rescaled $p-t$ graph) for RN-AdS black hole. Line 1 is from Eq.(24) and Eq.(40) and Eq.(52) which displays an overlap. The former two are obtained by using the equal area law in $T-S$ plane and $P-V$ plane, the later one is obtained by analyzing the characteristic swallow tail behavior of the Gibbs free energy. Line 2 is from Eq.(51) using the equal area law in $P-\nu$ plane.}  \label{tpt}
\end{center}
\end{figure}

We note that in Ref.\cite{2014shaowen}, the authors obtained a numerical result of the phase transition coexistence line by using the second method (analyzing the characteristic swallow tail behavior of the Gibbs free energy). Their fitting formula is (Eq.(45) in their paper)
\begin{eqnarray}
\tilde{P}&=&0.666902\tilde{T}^{2}+0.175830\tilde{T}^{3}+0.127273\tilde{T}^{4}\nonumber\\
&\,&-0.230638\tilde{T}^{5}+0.795846\tilde{T}^{6}-1.36972\tilde{T}^{7}\nonumber\\
&\,&+1.47494\tilde{T}^{8}-0.867209\tilde{T}^{9}+0.226773\tilde{T}^{10},\nonumber\\
&\,&\,\tilde{T}\in(0,1).
\end{eqnarray}
The fitting formula displays an overlap with our analytical results Eq.(24) and Eq.(40) in the $p-t$ graph Fig.\ref{tpt}. This also indicates that the phase transition coexistence line Eq.(51) is wrong.

That is why we emphasize that the equal area law can't be used in the $P-\nu$ plane. The equal area law comes from the first law and the first law comes from the Smarr relation. So the reason originates from the Smarr relation for the RN-AdS black hole. When the internal energy is treated as the enthalpy ($M\equiv{H}$) of the system, the Smarr relation is
\begin{equation}
H=2TS-2PV+Q\phi.
\end{equation}
Differentiating it, we get the first law
\begin{equation}
dH=TdS+VdP+\phi{d}Q.
\end{equation}
Then the Gibbs free energy can be obtained as
\begin{equation}
dG=-SdT+VdP+\phi{d}Q.
\end{equation}
During the phase transition, the chemical potential is identical between the two phases. So the Gibbs free energy is unchanged that means
\begin{equation}
-SdT+VdP+\phi{d}Q=0.
\end{equation}
Fixing $T$ and $Q$, we can get
\begin{equation}
\int_{P_{g}}^{P_{l}}VdP=0.
\end{equation}
This gives the equal area law
\begin{equation}
P(V_{g}-V_{l})=\int_{V_{l}}^{V_{g}}P(V)dV.
\end{equation}
By fixing $P$ and $Q$, or $T$ and $P$, we get the other two equal area laws in $T-S$ and $Q-\phi$ planes \cite{2013arXiv13053379S,2013arXiv1310.2186S}. The equal area law is right only for special planes rather than any kind of oscillating lines ( such as the $T-r$ plane in Fig.1 or the $P-\nu$ plane ). For the $P-V$ plane, the $V$ is the volume of the black hole rather than the specific volume $\nu$. That is why the equal area law can't be used in the $P-\nu$ plane. So far, we have explicitly checked that the equal area law can't be used in the $P-\nu$ plane by comparing the $p-t$ phase transition graphs obtained by using the equal area law in $T-S$, $P-V$, $P-\nu$ planes and by analyzing the characteristic swallow tail behavior of the Gibbs free energy, and it is also checked from the Smarr equation.

\section{Discussion and conclusion}
\label{conclusion}

 The equal area law is investigated for black hole phase transition. For the especial case RN-AdS black hole, we argued that the equal area law can be only used in the $P-V$, $T-S$ and $\phi-Q$ planes. Though in the $T-r$ or $P-\nu$ plane there is an oscillating behavior below the critical point, the equal area isobar can't be used to replace the oscillating part. To address this argument, first of all, we suppose that the equal area law holds for any state graph which possesses an oscillating behavior. Then we have obtained the phase transition points (an analytical relation between T,P,Q) by using the equal area law in $T-S$, $P-V$, $P-\nu$ planes. The result shows that the phase diagrams obtained from $T-S$, $P-V$ planes are identical but they are different from the one obtained from $P-\nu$ plane. There should be only one phase diagram for a thermodynamical system, so the phase diagram obtained from $P-\nu$ plane is wrong. We have also made a comparison of our results with the fitting formula Eq.(52) obtained in Ref.\cite{2014shaowen} which indicates the phase diagram obtained from $P-\nu$ plane is wrong. To further understand why the equal area law can't be used in these planes, we traced back to the derivation of the equal area law and found out that the Smarr relation or the first law which guarantees the equal area law can only be used in the $P-V$, $T-S$ and $\phi-Q$ planes.

\begin{figure}
\begin{center}
\includegraphics[width=0.50\textwidth]{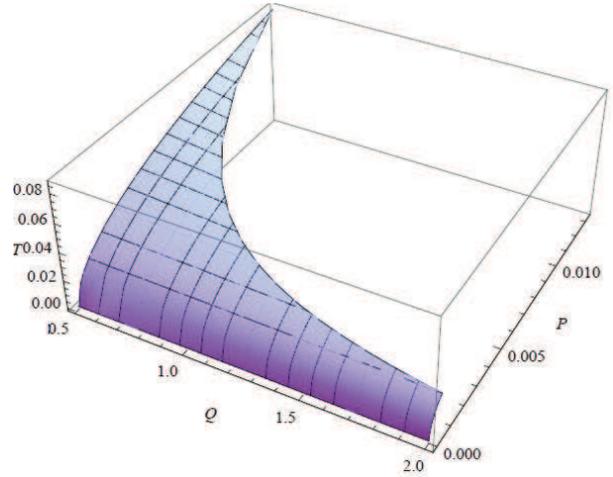}
\caption{The phase transition coexistence surface ( $T-P-Q$ graph) for RN-AdS black hole. The phase transition line of $P-T$ decrease as the charge $Q$ increase, the pressure $P$ increase as the temperature T increase}  \label{ttpq}
\end{center}
\end{figure}

In subsection \ref{ts} and \ref{pv}, we get the analytical phase transition relation $T-P-Q$. With this analytical phase transition relation, it is convenient to analyze the phenomena near the critical point. The graph is plotted in Fig.\ref{ttpq}.
We see that the RN-AdS black hole's $P-T$ coexistence line for fixed charge $Q$ is just the same with the VDW case: as the temperature $T$ increases during the smaller $T$ region, the corresponding pressure $P$ increases very slowly, while during the bigger $T$ region, $P$ increases quickly as $T$ increases. The critical points are very different for different $Q$, the critical point quickly decreases as the $Q$ increases. This means that the condition for a highly charged RN-AdS black hole to transit from its ``gas" phase to ``liquid" phase is very difficult to reach which requires very low temperature.

 The RN-AdS black hole system is a reminiscence of VDW system in many aspects except specific volume. For the VDW case, the equal area law can be used in $P-\nu$ plane. The specific volume : $\nu=\frac{V}{Nm}\sim\frac{V}{N}$, here $N$ stands for the molecule number in $V$ and $m$ stands for the molecule mass. The molecule mass is a constant, so the specific volume can stand for one molecule's volume. This means the specific volume gives us the microscopic information of the VDW system. We may expect that the specific volume of the black hole can give us some microscopic information too. The different specific volume between a RN-AdS black hole system and a VDW system may probably means the different microscopic structures.

\begin{acknowledgements}

This work is supported by the National Natural Science Foundation of China (Grant Nos. 11235003, 11175019, 11178007)

\end{acknowledgements}

\bibliographystyle{spphys}
\bibliography{construction}

\end{document}